\documentclass[a4paper,11pt,journal]{article}

\usepackage[title,titletoc,toc]{appendix}
\usepackage{graphicx}
\usepackage{latexsym}
\usepackage{amsmath}
\usepackage{hyperref}
\usepackage{subfig}
\usepackage{subfloat}
\usepackage[hypcap=true]{caption}
\usepackage{enumitem}
\DeclareGraphicsExtensions{.pdf,.png,.jpg}

\begin{document}

\title{Crunch-in regime - Non-linearly driven hollow-channel plasma}

\author{Aakash A. Sahai  \thanks{a.sahai@imperial.ac.uk},
		\\ John Adams Institute for Accelerator Science, Physics Department, \\
		Blackett Laboratory, Imperial College London, SW7 2AZ, UK }

\maketitle

% ABSTRACT
\begin{abstract}
Plasma wakefields driven inside a hollow-channel plasma are significantly different from those driven in a homogeneous plasma. This work investigates the scaling laws of the accelerating and focusing fields in the ``crunch-in'' regime \cite{phd-thesis}\cite{positron-nonlinear-wake}. This regime is excited due to the collapse of the electron-rings from the channel walls onto the propagation axis of the energy-source, in its wake. This regime is thus the non-linearly driven hollow channel, since the electron-ring displacement is of the order of the channel radius \cite{positron-nonlinear-wake}\cite{nonlinear-ion-wake}. We present the properties of the coherent structures in the ``crunch-in'' regime where the channel radius is matched to the beam properties such that channel-edge to on-axis collapse time has a direct correspondence to the energy source intensity. We also investigate the physical mechanisms that underlie the ``crunch-in'' wakefields by tuning the channel radius. Using a theoretical framework and results from PIC simulations the possible applications of the ``crunch-in'' regime for acceleration of positron beams with collider-scale parameters is presented.
\end{abstract}

%%%%%%%%%%%%%%%%%%%%%%%%%%%%%%%%%%%%%%%%%%%%%%%%%%%%%%%%%%%%%%%%%%%%%
% SECTION - Introduction
%%%%%%%%%%%%%%%%%%%%%%%%%%%%%%%%%%%%%%%%%%%%%%%%%%%%%%%%%%%%%%%%%%%%%
% INTRODUCTION
\section{Introduction}

Hollow-channel plasma wakefield are excited due to the dynamics of the channel-wall electrons of a certain density, driven by an energy-source (a beam of photons or particles) propagating in a very low density region such as vacuum, unionized gas or very-low density plasma. Hollow-channel wakefields are thus driven due to the dynamics of the channel-wall electrons exciting electron-ion charge-separation.

This work \cite{phd-thesis}\cite{positron-nonlinear-wake} counters the previously established conclusions about hollow-channel wakefields, that:
\begin{enumerate}[nolistsep,label=(\roman*)]
\item there are zero focusing fields on the axis of a hollow-channel
\item the accelerating fields along the channel axis are purely electromagnetic, because they are essentially the leakage (fringe) fields of the charge-separation excited in the channel-walls. 
\end{enumerate}

This work establishes that in the ``crunch-in'' regime irrespective of the driving energy-source, electrostatic fields are excited through density compression of the wall-electrons, that undergo a crunching-in process from the walls onto the channel axis. It is also shown that these fields approach the wave-breaking limit of the channel-wall electron density when the channel radius is matched to the energy-source. 

Hollow-channels with skin-depth scale tunable radius, that enable the ``crunch-in'' regime with accessible acceleration lengths over several 10s of centimeter length have been proposed using an {\it ion-wake} \cite{nonlinear-ion-wake}.

\begin{figure}[!htb]
   \centering
   \includegraphics*[width=4.5in]{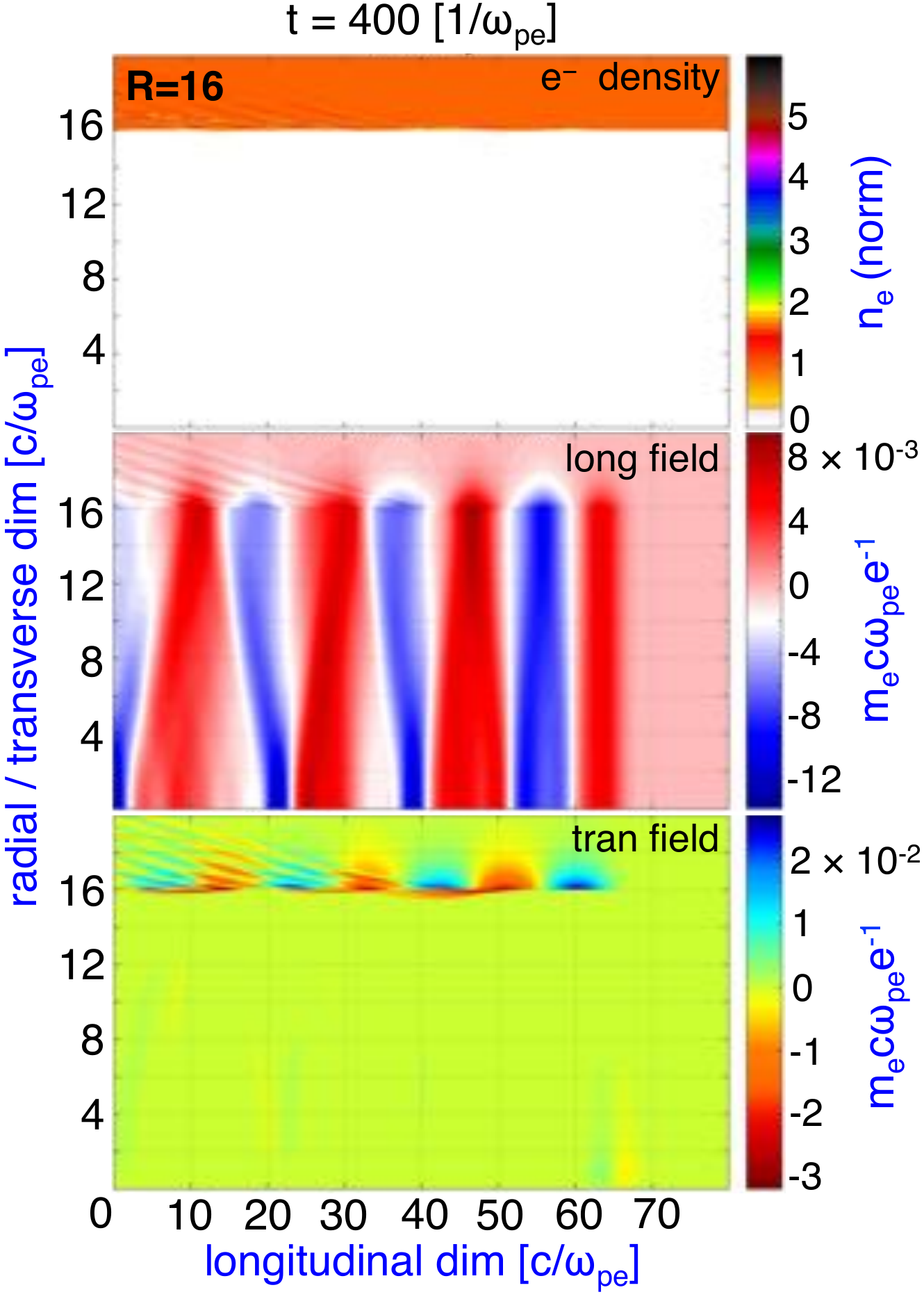}
   \caption{ {\bf $e^-$-beam driven linear Hollow-channel wakefields} The $e^-$ density, longitudinal and transverse plasma field in cylindrical geometry of a linearly-driven hollow-channel, $R=16$. The on-axis focusing field are nearly zero. The longitudinal-fields are excited by the leakage of the wall charge-separation fields into the center of the channel and are of the order of $10^{-2}~m_ec\omega_{pe}e^{-1}$. ($\gamma_b=10^3$, $\sigma_r = 0.5c/\omega_{pe}$, $\sigma_z = 1.5c/\omega_{pe}$, $n_b=1.5n_0$).}
   \label{Fig1:lin-hollow-e-beam}
\end{figure}

The distinction between previously explored hollow-channel regime and the ``crunch-in'' regime can be inferred by comparing electron-beam driven simulation snapshots ($e^-$ density, plasma acc. field and focusing field) in Fig.\ref{Fig1:lin-hollow-e-beam} (earlier hollow-channel regime) and Fig.\ref{Fig2:nonlin-hollow-e-beam} (``crunch-in'' regime).

%%%%%%%%%%%%%%%%%%%%%%%%%%%%%%%%%%%%%%%%%
% FIGURE - LINEAR regime Hollow-channel 
%%%%%%%%%%%%%%%%%%%%%%%%%%%%%%%%%%%%%%%%%
% Homogenous plasma - positron-beam excited wakefields

It is known that the properties of hollow-channel wakefields are dictated by 2 independent parameters:
\begin{enumerate}[noitemsep,nolistsep]
\item hollow-channel radius, $r_{\rm{ch}}$
\item channel-wall plasma-electron density, $n_0$ or the plasma-electron frequency, $\omega_{\rm{pe}}=\sqrt{4\pi n_0 e^2/m_e}$.
\end{enumerate}

However, in this work it is shown that the energy-source intensity determines the properties of the hollow-channel wakefields in the ``crunch-in'' regime. Intensity is represented with $a_0$, the normalized vector potential for lasers and with $n_b$, the beam density for particle-beams.

%%%%%%%%%%%%%%%%%%%%%%%%%%%%%%%%%%%%%%%%%%%%%%%%%%%%%%%%%%%%%%%%%%%%%
% SECTION - Accessing the Crunch-in regime
%%%%%%%%%%%%%%%%%%%%%%%%%%%%%%%%%%%%%%%%%%%%%%%%%%%%%%%%%%%%%%%%%%%%%
% ACCESSING THE CRUNCH-IN REGIME
\section{Accessing the crunch-in regime}

The previously analyzed linear-regime of hollow-channel wakefields operates under the assumption in eq.\ref{eq:linear-hollow-channel-condition} for photon and particle-beam energy sources respectively:
% equation for LINEAR HOLLOW CHANNEL CONDITION
\begin{equation}
\begin{aligned}
w_0 \simeq r_{ch}: ~ \frac{e}{m_ec\omega_{pe}} ~ \frac{e}{r_er_{\rm{ch}}}\frac{a_0^2}{2\gamma_e} & \ll 1 \\
\frac{e}{m_ec\omega_{pe}} ~ \frac{en_b}{r_{\rm{ch}}^2} \left(\frac{\pi}{2}\right)^{3/2} ~ \sigma_r^2 \sigma_z& \ll 1
\end{aligned}
\label{eq:linear-hollow-channel-condition}
\end{equation}
%\begin{equation}
%\begin{aligned}
%\frac{1}{R}\frac{a_0^2}{2\gamma_e} & \ll 1 \\
%\frac{1}{8}\sqrt{\frac{\pi}{2}} ~ \frac{1}{R^2} ~ \frac{n_b}{n_0} \left(\frac{\sigma_r}{c/\omega_{pe}}\right)^2\frac{\sigma_z}{c/\omega_{pe}} & \ll 1
%\end{aligned}
%\label{eq:linear-hollow-channel-condition}
%\end{equation}
\noindent where, $w_0$ is the laser focal spot-size, $\sigma_r$ is the bunch spot-size and $\sigma_z$ the bunch length. Note, the choice of $\sigma_r$ is not dictated by the channel radius, $r_{ch}$, unlike $w_0$ for lasers. 

%%%%%%%%%%%%%%%%%%%%%%%%%%%%%%%%%%%%%%%%%%%%
% FIGURE - NON-LINEAR regime Hollow-channel 
%%%%%%%%%%%%%%%%%%%%%%%%%%%%%%%%%%%%%%%%%%%%
% Homogenous plasma - positron-beam excited wakefields
\begin{figure}[!htb]
   \centering
   \includegraphics*[width=4.5in]{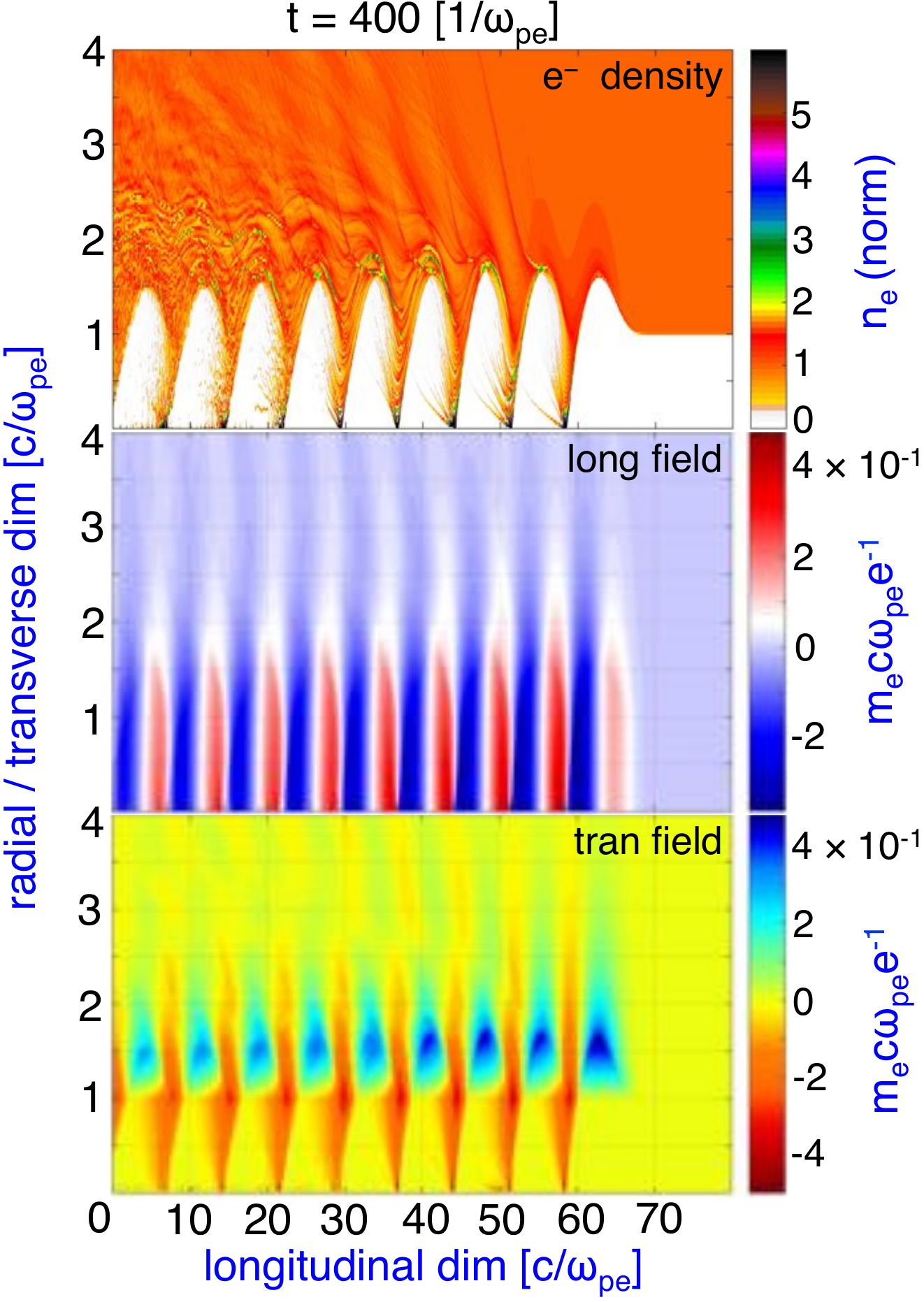}
   \caption{ {\bf $e^-$-beam driven matched ``crunch-in'' wakefields.} The $e^-$ density, acc. and transverse plasma field for the ``crunch-in'' regime in cylindrical geometry, $R=1$. Notably, there is $e^-$ density compression along the axis of the channel. Additionally, the acc. and focusing fields are of the order of the wave-breaking fields of the channel-wall electron density. (same $e^-$-beam parameters as in Fig.\ref{Fig1:lin-hollow-e-beam}) }
   \label{Fig2:nonlin-hollow-e-beam}
\end{figure}

The ``crunch-in'' regime requires two conditions which dictate the radial dynamics of channel-edge electrons:
\begin{enumerate}[nolistsep,label=(\alph*)]
\item the {\it channel radius is small enough} to allow on-axis collapse of the channel-edge electrons within a spatial-scale ($\delta r$) of the order of a plasma ($n_0$) wavelength, $2\pi c/\omega_{pe}$ 
\item the energy-source is {\it intense enough} to drive the electrons to velocities ($\delta r / v_e$) that allow oscillations over the time-scales of the order of a plasma ($n_0$) period, $2\pi\omega_{pe}^{-1}$
\end{enumerate}

Thus, the condition for the excitation of the ``crunch-in'' regime for laser and particle-beam respectively, are in eq.\ref{eq:nonlinear-hollow-channel-condition}:
% equation for ``CRUNCH-IN'' HOLLOW CHANNEL CONDITION - CHANNEL-RADIUS
\begin{equation}
\begin{aligned}
R_N & \simeq \frac{a_0^2}{2\gamma_e} \quad (k_pw_0 \simeq R)\\
R_N & \simeq \frac{n_b}{n_e} ~ \frac{\sigma_r}{c/\omega_{\rm{pe}}}\sqrt{ \frac{\sigma_z}{c/\omega_{\rm{pe}}} } ~ \frac{1}{2\sqrt{2}}\left(\frac{\pi}{2}\right)^{1/4}
\end{aligned}
\label{eq:nonlinear-hollow-channel-condition}
\end{equation}
\noindent where, $R_N=r_{\rm{ch}} \left(c/\omega_{\rm{pe}}\right)^{-1}$ leads to wave-breaking fields.

This work \cite{phd-thesis}\cite{positron-nonlinear-wake} also shows that the eq.\ref{eq:nonlinear-hollow-channel-condition} defines the relation of the wakefield amplitude to $R_N$ and energy-source intensity ($a_0$ and $n_b/\sigma_r/\sigma_z$), for approaching non-linearity. For instance using the $e^-$ beam parameters presented in Fig.\ref{Fig1:lin-hollow-e-beam} in eq.\ref{eq:nonlinear-hollow-channel-condition}, the value of $R_N = 0.36$. 

However, there exists a {\it matching condition} to excite optimal ``crunch-in'' wakefields \cite{positron-nonlinear-wake}. It is also not necessarily optimal to set the channel radius to access the wave-breaking limit when considering these wakefields for acceleration.

By comparing the accelerating fields of the ``crunch-in'' regime (in Fig.\ref{Fig2:nonlin-hollow-e-beam}) to that of the linear regime (in Fig.\ref{eq:linear-hollow-channel-condition}) it is seen that there is a 2 orders of magnitude increase in amplitude. This is attributed to the origin of these higher fields being in the on-axis density compression due to the crunching-in electrons from the channel-walls.

Similarly, comparing the traverse / focusing force we find that whereas in the linear regime there are nearly zero focusing force, in the ``crunch-in'' regime there is a significant focusing force of the same order as the wave-breaking field of the channel-wall density.

Another important point to note is that the phase of the focusing fields in Fig.\ref{Fig2:nonlin-hollow-e-beam} are negative for positive radius, which means it is attractive for a positively charged particle propagating at the same speed as the wake. Thus, these strong focusing fields on-axis are useful for transporting positron beams. It is also important to note that the accelerating phase (positive electric field - in red) of the longitudinal wakefields spatially overlaps with the focusing phase of transverse wakefields, ideal for loading positron witness bunches.

%%%%%%%%%%%%%%%%%%%%%%%%%%%%%%%%%%%%%%%%%%%%%%%%%%%%%%%%%%%%%%%%%%%%%
% SECTION - CRUNCH-IN DYNAMICS
%%%%%%%%%%%%%%%%%%%%%%%%%%%%%%%%%%%%%%%%%%%%%%%%%%%%%%%%%%%%%%%%%%%%%
% DYNAMICS OF THE WALL ELECTRONS
\section{Dynamics of the wall electrons}

The magnitude of the radial perturbation of the electron rings that are initially located at the channel wall, is defined as, 
% electron ring radial perturbation magnitude
$$ \delta r(\xi) = r(\xi) - r_{ch} $$ 
(where $\xi=v_{es}t-z$ and $v_{es}$ is the energy-source velocity) dictates whether the fields are purely electromagnetic (linear regime) or predominantly electrostatic (nonlinear regime). The radial electron perturbation relative to the channel radius is a key metric to understand the properties of the wakefields. 

A relativistic beam of particles has zero focusing-fields on-axis only if the channel-wall electrons are perturbed weakly. The reason for this is that their radial perturbation is much smaller than the channel radius, 
% radial perturbation SMALLER than channel radius
$$\delta r \ll r_{ch}$$ 
and thus the electrostatic fields of charge-separation in channel walls only leak as fringe fields into the hollow channel. Thus, here it is only the electron current driven in the channel walls which results in the excitation of the wakefields.

However, as it has been shown in \cite{phd-thesis}\cite{positron-nonlinear-wake}, that if the channel radius is a few skin-depths, then,
% radial perturbation GREATER than channel radius
$$\delta r \geq r_{ch}$$ 
as a result of which the electron rings from the channel wall collapse to the axis along which the center of the driving energy source propagates. 

%%%%%%%%%%%%%%%%%%%%%%%%%%%%%%%%%%%%%%%%%%%%%
% FIGURE - p2x1 - Linear vs Crunch-In regime 
%%%%%%%%%%%%%%%%%%%%%%%%%%%%%%%%%%%%%%%%%%%%%
% Homogenous plasma - positron-beam excited wakefields
\begin{figure}[!htb]
   \centering
   \includegraphics*[width=4.5in]{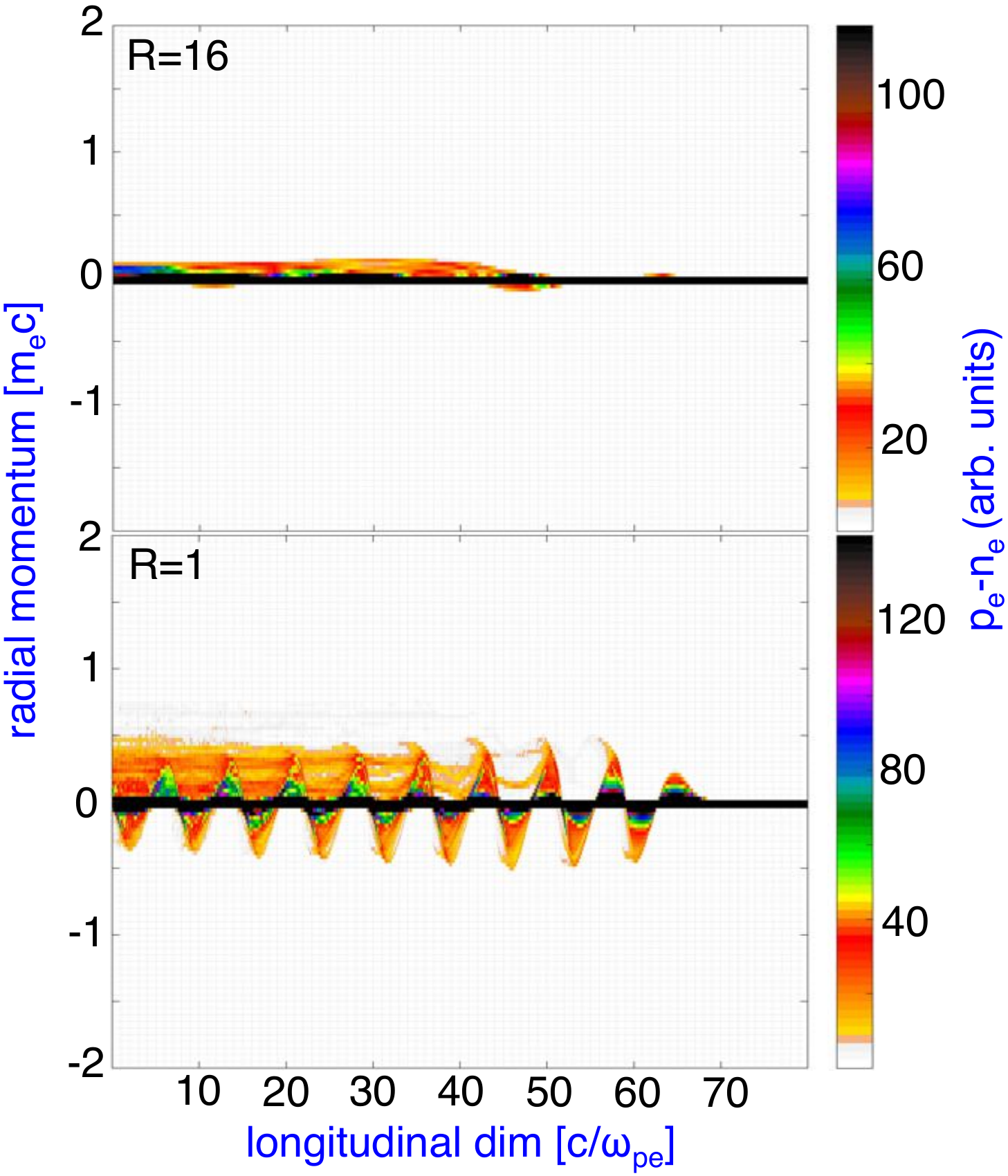}
   \caption{ {\bf Radial momentum Phase -space - linear vs Crunch-in regime.} Here the radial $e^-$ momentum excited by the $e^-$ beam is compared for the two cases above. For linear regime corresponding to Fig.\ref{Fig1:lin-hollow-e-beam} the maximum radial momentum is $p_r \leq 0.1m_ec$, whereas in the crunch-in regime (corresponding to Fig.\ref{Fig2:nonlin-hollow-e-beam}), $p_r \simeq 0.5m_ec$.}
   \label{Fig3:Lin-VS-NonLin-p2x1}
\end{figure}

The magnitude of the radial perturbations is dictated by the force of the energy-source on the electrons at the channel wall (setting $v_{es}\simeq c$ and ignoring $\delta r^2$ term),
% EQUATION OF MOTION - electron rings
\begin{equation}
\begin{aligned}
\frac{\partial^2}{\partial\xi^2} ~ r(\xi) + \frac{1}{2}\frac{\omega_{pe}^2}{c^2} ~ \left(1-2\frac{r_{ch}}{r(\xi)}\right) ~ r(\xi) & = \frac{F_{r=r_{ch}}(\xi)}{m_ec^2} \\
\frac{\partial}{\partial\xi} ~ \frac{p_r(\xi)}{m_ec} + \frac{1}{2} \frac{\omega_{pe}}{c} ~ \frac{(r-2r_{ch})}{c/\omega_{pe}} & = \frac{F_{r=r_{ch}}(\xi)}{m_ec^2}
\end{aligned}
\label{eq:ring-equation-of-motion}
\end{equation}

The eq.\ref{eq:ring-equation-of-motion} can be compared to the equation for the collapse of the electron rings in a homogeneous plasma when driven by a positron beam.
$$\frac{\operatorname d^2}{\operatorname d\xi^2}r = - \frac{1}{r} n_{b}(\xi)\frac{\pi}{2} \sigma_r^2(\xi)$$ which can be solved for the collapse time (that is when $1/r \rightarrow \infty$), $t_{coll}\omega_{pe}=\sqrt{2} \frac{ r_{ch}/\sigma_r }{ \sqrt{n_b/n_0} }$. Using this to calculate the collapse time for the case in Fig.\ref{Fig2:nonlin-hollow-e-beam}, we find that $t_{coll} = 2.3\omega_{pe}^{-1}$. This can be compared to the collapse time for an electron beam which is about $10\omega_{pe}^{-1}$ in Fig.\ref{Fig2:nonlin-hollow-e-beam}. The collapse time is thus very different depending upon the nature of the driver and whether the plasma is homogeneous or hollow.

The radial momentum phase-spaces for the two cases compared in Fig.\ref{Fig1:lin-hollow-e-beam} and Fig.\ref{Fig2:nonlin-hollow-e-beam}, are presented in Fig.\ref{Fig3:Lin-VS-NonLin-p2x1}.

It can be seen that the momentum in the ``crunch-in'' regime is significantly higher than in the linear hollow-channel regime. Further, in the ``crunch-in'' regime the momentum phase-space has the characteristic oscillation wavelength of $2\pi c/\omega_{pe}$.

In simulations that are not presented here, it is observed that similar channel radius matching condition is obtained for a laser and a positron beam based driving energy-source.

\section{Discussion}
We have shown that ``crunch-in'' or the non-linearly driven hollow-channel regime is accessible and optimal under a right matching condition between the energy source intensity and the channel radius.

It is also shown that in the ``crunch-in'' regime the accelerating and the focusing fields are significantly higher and comparable to the wave-breaking fields of the channel-walls. There is also a significant focusing field which is known to be zero on-axis in previous studies on hollow-channels.

In continuation of investigations into this work we intend to continue exploring the channel wall electron dynamics to establish the optimal condition for the ``crunch-in'' regime with varying driver parameters. The work on further investigating positron acceleration and collider-scale electron acceleration in this regime will be further continued.

\section{Acknowledgment}
Work supported by the John Adams Institute for Accelerator Sciences \& Department of Physics, Imperial College London, London SW7 2AZ, United Kingdom. I would also like to acknowledge the HPC facilities, HPC team and Mr. Simon Burbidge for providing large-scale computing facilities at Imperial.

\newpage
%%%%%%%%%%%%%%%%%%%%%%%%%%%%%%%%%%
% REFERENCES
%%%%%%%%%%%%%%%%%%%%%%%%%%%%%%%%%%

\end{document}